# Combinatorial Printing of Functionally Graded Solid-State Electrolyte for High-Voltage Lithium Metal Batteries

*Qiang Jiang, Stephanie Atampugre, Yipu Du, Lingyu Yang, Jennifer L. Schaefer *,*

*Yanliang Zhang **

Q. Jiang, S. Atampugre, Y. Du, Prof. Y. Zhang

Department of Aerospace and Mechanical Engineering,

University of Notre Dame,

Notre Dame, IN 46556, USA

E-mail: yzhang45@nd.edu

L. Yang, Prof. J. L. Schaefer

Department of Chemical and Biomolecular Engineering,

University of Notre Dame,

Notre Dame, IN 46556, USA

E-mail: Jennifer.L.Schaefer.43@nd.edu

Abstract

Heterogeneous multilayered solid-state electrolyte (HMSSE) has been widely explored for their broadened working voltage range and compatibility with electrodes. However, due to the limitations of traditional manufacturing methods such as casting, the interface between electrolyte layers in HMSSE can decrease the ionic conductivity severely. Here, a novel combinatory aerosol jet printing (CAJP) is introduced to fabricate functionally graded solid-state electrolyte (FGSSE) without sharp interface. Owing to the CAJP's unique ability (in-situ mixing and instantaneous tuning of the mixing ratio), FGSSE with smooth microscale compositional gradation is achieved. Electrochemical tests show that FGSSE has excellent oxidative stability exceeding 5.5 V and improved conductivity (>7 times of an analogous HMSSE). By decoupling the total resistance, we show that the resistance from the electrolyte/electrolyte interface of HMSSE is 5.7 times of the total resistance of FGSSE. The Li/FGSSE/NCM622 cell can be stably run for more than 200 cycles along with improved rate performance.

**1. Introduction**

High-voltage solid-state lithium metal batteries (SSLMBs) are advanced energy storage devices which have attracted extensive interests owing to the enhanced energy density and safety. It can meet the requirement of the electrification of transportation and the storage of intermittently produced solar and wind energy.[1–6] Various types of lithium-conducting solid electrolytes have been explored as solid-state electrolytes (SSEs) in SSLMBs.[7–11] Nonetheless, SSEs, whether made of ceramics, polymers, or composite materials, face challenges when it comes to facilitating the chemistries of both the anode and cathode. Some SSEs exhibit good stability when paired with a reducing Li metal anode, but their limited resistance to oxidation makes them incompatible with high-voltage cathodes, thereby restricting the energy density of SSLMBs. On the other hand, SSEs that are compatible with high-voltage cathodes often suffer from instability when used with a Li metal anode, significantly limiting their versatility and practicality. Considering that each SSE has its own advantages and drawbacks, it becomes a formidable task to identify a single SSE with the ability to withstand both reduction and oxidation simultaneously.

Instead of searching for a single SSE with a chemical composition capable of handling both reduction and oxidation simultaneously, heterogeneous multilayered solid-state electrolyte

(HMSSE) strategy is introduced to solve this dilemma.[12–18] For instance, Goodenough et al. first applied this strategy via a dual-layered solid electrolyte consisting of polyethylene oxide (PEO) polymer contacting the lithium-metal anode and a poly(N-methyl-malonic amide) (PMA) contacting the cathode.[19] Lu et al. constructed a dual-layered SSE by leveraging the oxidation resistance of poly(acrylonitrile) (PAN) and reduction compatibility of poly(vinylidene fluoride) (PVDF) layer.[16] Sandwich structure was also proposed in addition to the bilayer structure. Guo et al. further extended the electrochemical range of SSEs to 0–5 V by implementing a complex triple-layer SSE approach, using oxidation-resistant PAN and reduction-tolerant polyethylene glycol diacrylate (PEGDA) layers in contact with high-voltage cathodes and Li metal anodes, respectively, while incorporating a flexible PAN@$Li_{1.4}Al_{0.4}Ge_{1.6}(PO_4)_3$ composite electrolyte as an intermediate layer to inhibit dendrite formation and ensure stable operation of high-voltage cathodes.[20] Although the HMSSE strategy can potentially broaden the working voltage windows, the newly introduced interface between electrolyte layers, which is unavoidable in the HMSSE prepared by traditional manufacturing methods such as casting, has raised new challenges when considering the ion transport in the electrolyte. The ion transport resistance at the interface can be much higher than that from bulk electrolyte, leading to the inferior performance of the HMSSE.[21–27] According to the in-depth electrochemical impedance spectroscopy (EIS) analysis and simulation results, Brandell et al. demonstrated that the interfacial resistance can be approximately 10-fold higher than that of the bulk ionic resistance.[21] Bouchet et al. deduced that the electrolyte/electrolyte interfacial resistance can be 100 times of that of the bulk electrolyte in the PEO-LATP multilayer model.[22] Thus, new manufacturing strategies are imperative to reduce the interfacial resistance and further improve the conductivity and the overall performance of the high-voltage SSLMBs.

Additive manufacturing (AM) has arisen as a versatile method for producing complex structures employing micro- and nanoscale building blocks.[28–30] The capability to combine the design freedom of additive manufacturing with precise control over material composition at the local level holds the potential for producing complex materials that are not attainable through traditional manufacturing methods.[31] Among the emerging AM approaches, aerosol jet printing has gained widespread attention due to its high resolution in material deposition and its wide applicability, encompassing materials such as polymers, ceramics, metals, semiconductors, adhesives, and biomaterials.[32–39]

Here, a novel combinatory aerosol jet printing (CAJP) method is introduced to fabricate functionally graded solid-state electrolyte (FGSSE). As shown in Figure 1a, oxidation-tolerant PAN and reduction-resistant PEO are separately dissolved and used as two inks in the CAJP. Lithium bis(trifluoromethane)sulfonimide (LiTFSI) salt and the nano-size $Li_{6.4}La_3Zr_{1.4}Ta_{0.6}O_{12}$ (LLZTO) particles as inorganic filler, which can promote the transport of $Li^+$, are also added into both inks.[40–42] In CAJP, two ultrasonic atomizers are used to convert liquid inks into aerosols containing microscale ink droplets. The two aerosolized ink streams are transported by $N_2$ carrier gas and then mixed within a single nozzle. The resulting mixture is directed and focused by a co-flowing sheath gas to achieve high spatial resolution before being deposited on a substrate. A unique advantage of aerosol-based ink deposition is the ability to realize in-situ mixing of multiple aerosolized inks and quickly change the mixing ratio due to the extremely low viscous drag of the aerosols compared with liquid or solid feedstock materials. Owing to the instantaneous mixing and tuning of the mixing ratio of two aerosols, the composition can be finely tuned along with the thickness direction of the printed materials to realize the FGSSE. This FGSSE displays an excellent oxidative stability exceeding 5.5 V and improved ionic conductivity (over 7 times of that of HMSSE). The high-voltage lithium metal cell with FGSSE can stably run for more than 200 cycles.

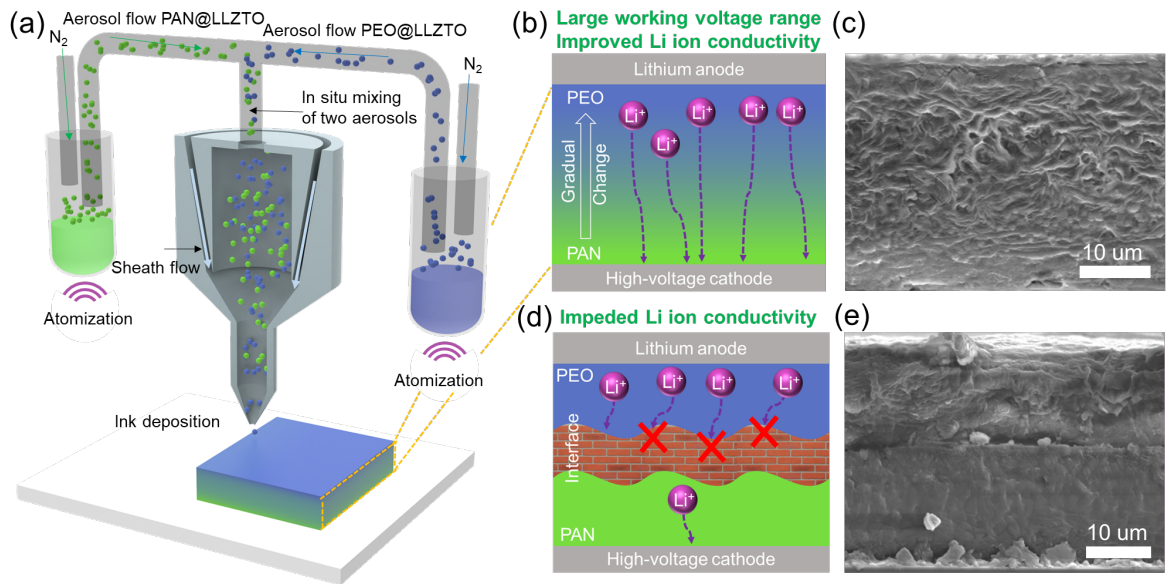

Figure 1 (a) Schematic illustration of the combinatorial printing of FGSSE using aerosolized inks. (b) Schematic illustration of the FGSSE with enhanced ion transport capability. (c) Cross-sectional SEM image of the FGSSE. (d) Schematic illustration of the HMSSE with impeded ion transport capability. (e) Cross-sectional SEM image of the HMSSE.

## 2. Results and Discussion

### 2.1. Combinatorial printing of electrolytes

To print the FGSSE with Z-axis gradient structure, 100% PAN@LLZTO ink was first deposited onto high-voltage cathode discs. As we continuously print the FGSSE with increasing thicknesses, the flow rate of PAN@LLZTO ink was gradually decreased from 100% to 0 and the flow rate of PEO@LLZTO ink was gradually increased from 0 to 100%. Finally, a gradient structure electrolyte with pure PAN@LLZTO at the bottom and pure PEO@LLZTO on the top was obtained. More details about the printing process can be found in the supporting information. As shown in Figure 1b and c, the pure PAN@LLZTO electrolyte contacts the high-voltage cathode which avoids the oxidation of PEO electrolyte when the batteries are run at high-voltage. And the pure PEO@LLZTO contacts the lithium metal which can separate the PAN from lithium metal avoiding the side reaction between PAN and the lithium metal. More importantly, the compositional ratio between PAN and PEO was gradually changed from 1:0 to 0:1 during the printing process along Z-axis of the FGSSE. The gradual and smooth compositional modulations avoid the sharp compositional changes and render the electrolyte merging very well between different layers of printing. In this FGSSE, the $Li^+$ can transfer efficiently across the whole electrolyte. In contrast, the HMSSE with pure PAN@LLZTO at the bottom and pure PEO@LLZTO on the top shows a sharp interface between the PAN@LLZTO and the PEO@LLZTO (Figure 1d and e), which acts as a barrier hindering the $Li^+$ transport across the electrolyte and reducing the $Li^+$ conductivity.

### 2.2. Characterizations of the printed electrolytes

To characterize the compositional distribution of PAN and PEO in FGSSE. EDS and Raman measurements were conducted to reveal the element and the organic functionality distributions across the FGSSE. Figure 2a shows N and O distributions along the cross-section of FGSSE film. At the bottom of the film, there is a dense layer of N signal indicating the bottom layer is composed of PAN. Right above the PAN layer, the N signal decreases, and the O signal increases gradually along the Z axis from the bottom to the top, which indicates the content of PAN decrease and the content of PEO increase gradually. This smooth compositional change can be attributed to the in-situ mixing and accurate modulation of the ink mixing ratio during deposition. At the top of the film, the O signal becomes more obvious demonstrating the PEO dominates the FGSSE. Few N signals are detected, this might be the N signal from high boiling point solvent (DMF and NMP) residue.[43] Figure 2b shows the homogeneous distribution of Zr element, indicating the good dispersion of the LLZTO nanoparticles within the polymer electrolyte. The confocal Raman microscope can distinguish different polymers and their spatial distribution by revealing the organic functionalities information. So, Raman spectra were collected at every 5 μm along the Z-axis of the electrolyte film to verify the spatial distribution of the polymers. In Figure 2c, the peak at 848 cm$^{-1}$ can be assigned as C-O stretching from PEO and the peak at 2244.7 cm$^{-1}$ can be assigned as C≡N peak from PAN, respectively.[44,45] The intensity of C≡N peak becomes weaker, and the C-O peak becomes stronger from the bottom to the top. Since the C≡N peak located at 2244.7 cm$^{-1}$ doesn't overlap with any other peaks, and therefore its intensity was used to estimate the relative content of PAN to understand the spatial distribution of PAN.[46] Figure 2d shows the normalized intensity of C≡N peak. The peak intensity decreases along the film thickness direction revealing the composition transition from PAN to PEO across the electrolyte film. The relative interaction between polymer, LiTFSI and LLZTO additives were also investigated by Raman, FTIR, and XRD (Figures S7-9).

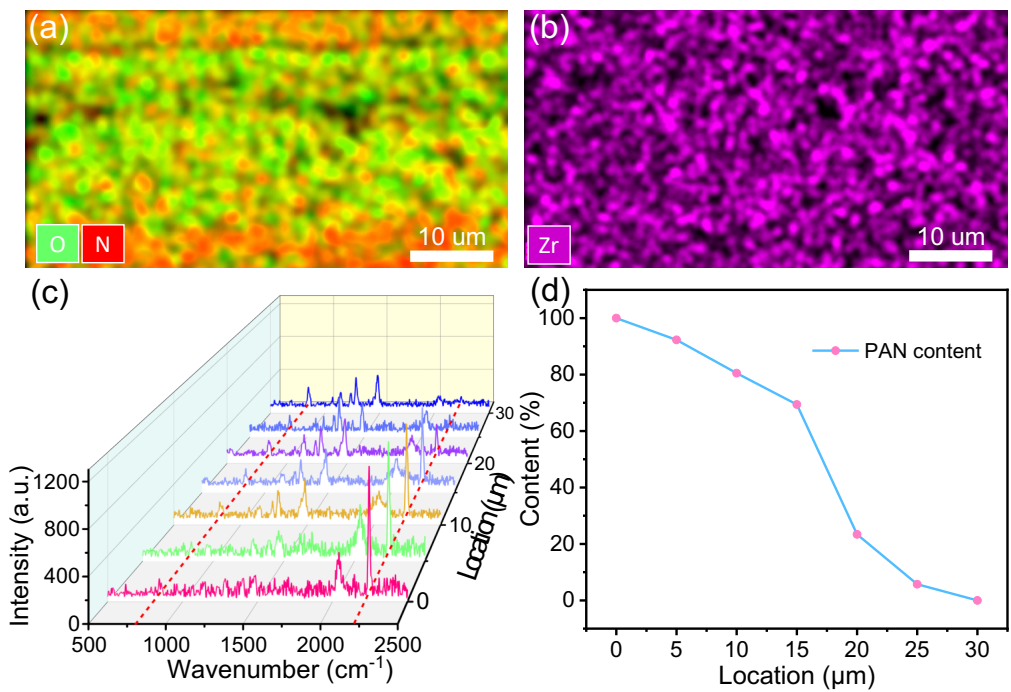

Figure 2 (a) EDS of N and O signals of FGSSE. (b) EDS of Zr signal of FGSSE. (c) Raman spectra at every 5 μm along the Z-axis of the FGSSE. (d) Normalized intensity of C≡N peak at different locations.

## 2.3. Electrochemical test of electrolytes

The ionic conductivity of the FGSSE and HMSSE was systematically studied by dielectric spectroscopy at different temperatures from 25 to 85°C. As shown in Figure 3a and Figure S10, the ionic conductivity increases with the increase of the temperature for both electrolytes. The conductivities at different temperatures are summarized and compared in Figure 3b. The FGSSE displays a conductivity of $2.0 \times 10^{-5}$ S cm$^{-1}$ at 25 °C which is over 7 times of that of HMSSE ($2.8 \times 10^{-6}$ S cm$^{-1}$ at 25 °C). This reveals that the FGSSE has an improved ion transport capability compared with the HMSSE. The activation energy ($Ea$) of conduction for each electrolyte is challenging to obtain because of the complex composition of electrolytes, rendering the use of a single model (Vogel−Tammann−Fulcher (VTF) equation or Arrhenius equation) inapplicable.[47] Figure 3c and Figure S11 show the EIS and DC polarization experimental results of symmetric lithium cells with FGSSE and HMSSE for Li$^+$ transference measurements. Based on these results and according to the Bruce–Vincent–Evans equation,[48–50] the Li$^+$ transference number ($t_{Li^+}$) of FGSSE is estimated to be 0.32 and $t_{Li^+}$ of the HMSSE is about 0.28. The electrochemical stability window is a critical property that determines whether the electrolyte is suitable for high-voltage Li metal batteries. The electrochemical window of the electrolyte was investigated by linear sweep voltammetry (LSV) test in Li/electrolyte/stainless steel (SS) cells at a scan rate of 0.2 mV s$^{-1}$. Adverse reactions start to happen for the PEO electrolyte when the voltage goes above 3.8 V, indicating its poor antioxidation performance (Figure S12). For FGSSE (Figure 3d), the excellent anti-oxidation capability is observed as with HMSSE (Figure S13). No obvious onset of current is observed until 5.5 V versus Li/Li$^+$, indicating its exceptional high-voltage stability.

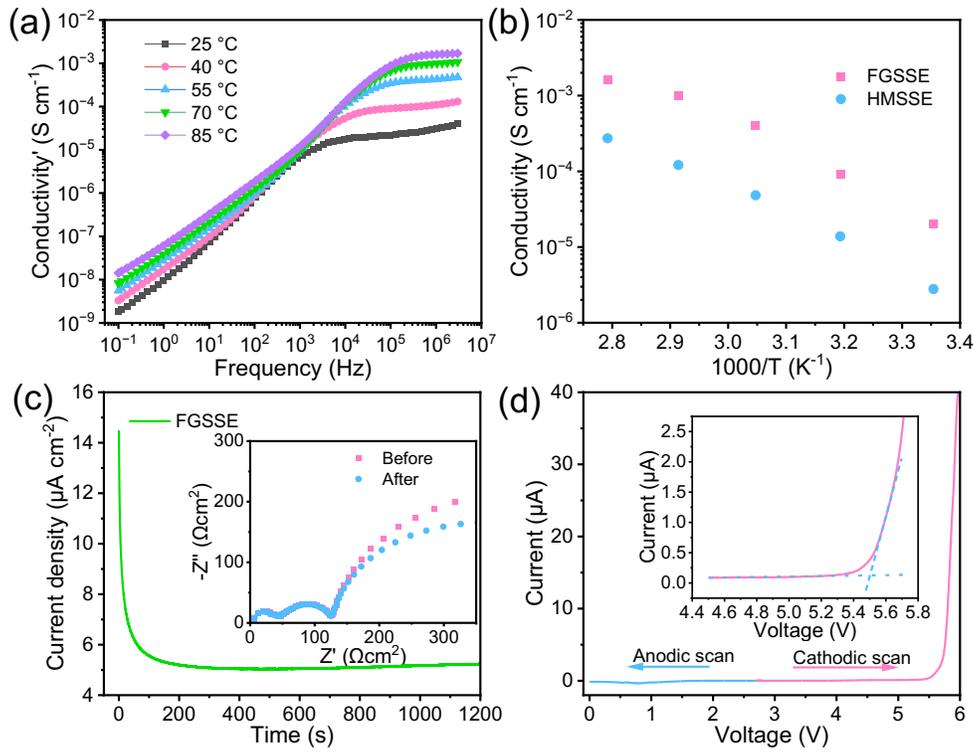

Figure 3 (a) Real conductivity as a function of frequency and temperature for FGSSE. (b) Temperature-dependent conductivity of FGSSE and HMSSE. (c) Impedance spectra and DC polarization curve of Li/FGSSE/Li cell for Li$^+$ transference number test. (d) Linear scan voltammetry of Li/FGSSE/SS. Inset is the zoom-in image of the onset of the oxidation process.

The performance of the printed electrolytes for energy storage application was evaluated by coin cell batteries with lithium anode and NCM622 cathode. Figure 4a presents the EIS spectra recorded at 60 °C for the cells Li/FGSSE/NCM622 and Li/HMSSE/NCM622. The spectra can be divided into high-frequency (HF, the first semicircle), medium-frequency (MF, the second semicircle) and low-frequency domains (LF, linear tail). And the HF, MF, and LF contributions are, respectively, assigned to the bulk electrolyte response, the interface response comprising both the polymer/polymer electrolyte interface and electrode/electrolyte interfaces, and the diffusion process.[24,52] The total polarization resistance ($R_p$) of the cell is the sum of bulk electrolyte (migration, $R_b$), interfacial charge transfer ($R_{ct}$), and diffusion resistance ($R_{dif}$), i.e., $R_p = R_b + R_{ct} + R_{dif}$. The $R_b$, $R_{ct}$, and $R_{dif}$ are well decoupled in frequencies, thus the $R_{ct}$ from different cells can be extracted from the EIS and compared directly. For cell with FGSSE, the $R_{ct}$ consists of Li/PEO and PAN/cathode interfaces. The $R_{ct}$ of cell with HMSSE is composed of Li/PEO, PAN/cathode, and the polymer/polymer electrolyte interfaces. According to the difference methodology,[22] the polymer/polymer interface resistance in the HMSSE can be deduced by subtracting the $R_{ct}$ of FGSSE from the $R_{ct}$ of HMSSE. To obtain the $R_{ct}$ from each electrolyte, the equivalent electrical circuit presented in Figure 4b is applied to fit the impedance spectra. The equivalent circuit is composed of the cable contribution [resistance ($R_c$) and inductance ($L_c$)] in series with the bulk electrolyte and charge transfer response (modeled by $R_b$//CPE$_b$ and $R_{ct}$//CPE$_{ct}$), and the Warburg impedance ($Z_w$). According to the fitting result, the $R_{ct}$ from HMSSE is 397.4 ± 4.2 Ωcm$^2$, while the $R_{ct}$ from FGSSE is 64.3 ± 0.4 Ωcm$^2$. Thus, polymer/polymer interface resistance in the HMSSE is calculated to be 333.1 ± 4.2 Ωcm$^2$, which is 5.7 times of the total resistance of FGSSE (58.6 Ωcm$^2$). This reveals that the significant charge transfer resistance arising from polymer/polymer interface can be dramatically reduced via our CAJP process.

Cyclic voltammetry (CV) measurement of the cell shows a typical NCM622 oxidization peak at 3.9 V and a reduction peak at 3.58 V (Figure 4c). In the first cycle, the oxidation peak shifts to 4.06 V because of the formation of cathode electrolyte interphase (CEI) during the first cycle. No other redox peak can be found, which also confirms the high stability of gradient electrolyte when it is paired with high-voltage cathode. Figure 4d shows the performance of the cells at various cycling rates from 0.2 to 5 C. The cell with FGSSE displays discharge capacities of 161, 153, 136, 110, and 44 mAh g$^{-1}$ at current rates of 0.2, 0.5, 1, 2, and 5 C respectively. In contrast, the cell with HMSSE displays discharge capacities of 160, 136, 103, 43, and 0.3 mAh g$^{-1}$ at corresponding rates. After the end of cycling at a high current rate of 5 C, the discharge capacity of FGSSE cell can be still boosted to 155 mAh g$^{-1}$ when the current rate returns to 0.2 C, reaching 96% of the initial discharge capacity at 0.2 C. While cells with different electrolytes exhibit similar capacity at relatively low currents, the cell with HMSSE displays significantly lower capacity once the current exceeds 1 C. Especially, when the current is increased to 5 C, the cell with HMSSE shows almost no capacity while the cell with FGSSE can still run and deliver a capacity of 44 mAh g$^{-1}$. This demonstrates the battery with FGSSE has much better rate performance.

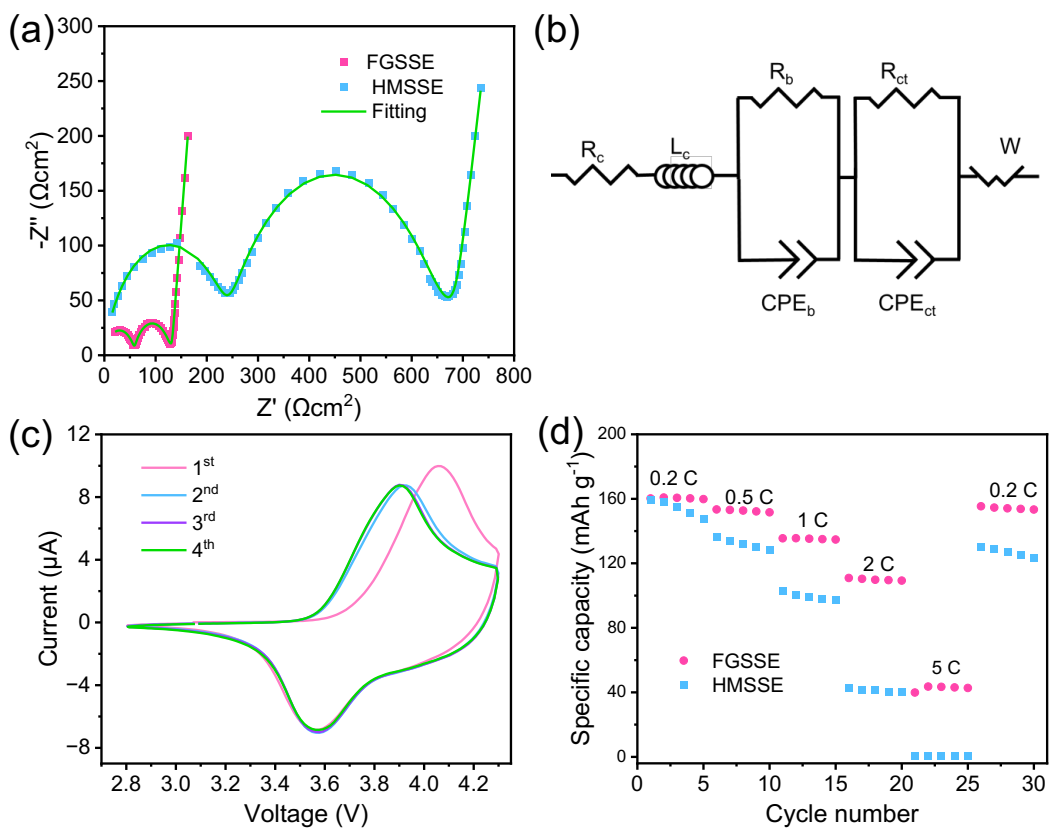

Figure 4 (a) Nyquist plots of the Li/FGSSE/NCM622 and Li/HMSSE/NCM622 cells. (b) The equivalent circuit model used to fit the EIS spectra. (c) CV curves of the Li/FGSSE/NCM622 cell. (d) Rate performance of the Li/FGSSE/NCM622 and Li/HMSSE/NCM622 cells. All measurements were conducted at 60 °C.

The long-term cycling performance of the cells is presented in Figure 5a in terms of the discharge capacity and Coulombic efficiency as a function of cycle number. The battery with FGSSE delivers a capacity of 142 mAh g$^{-1}$ after a few cycles of activation and maintains a capacity of 97 mAh g$^{-1}$ after 200 cycles, while the capacity of cell with HMSSE is about 113 mAh g$^{-1}$ after activation and just 40 mAh g$^{-1}$ after 200 cycles. The cell with FGSSE upholds a stable Coulombic efficiency throughout the cycling test at a level of 98.6−100% after the initial activation. After cycling, the cells were disassembled to observe the surface morphology evolution of Li metal anodes. A flat and smooth surface without porous or dendrite structure can be clearly observed for the Li metal from the cell with FGSSE after 200 cycles (Figure 5c). In sharp contrast, cracks and dendrite Li apparently existed in Li metal anode from the cell with HMSSE after cycling (Figure

5d). The growth of Li dendrites can be ascribed to the severe polarization due to the limited ionic diffusion and low $t_{Li^+}$ caused by the solid-state electrolyte/electrolyte interface.[51,53–57]

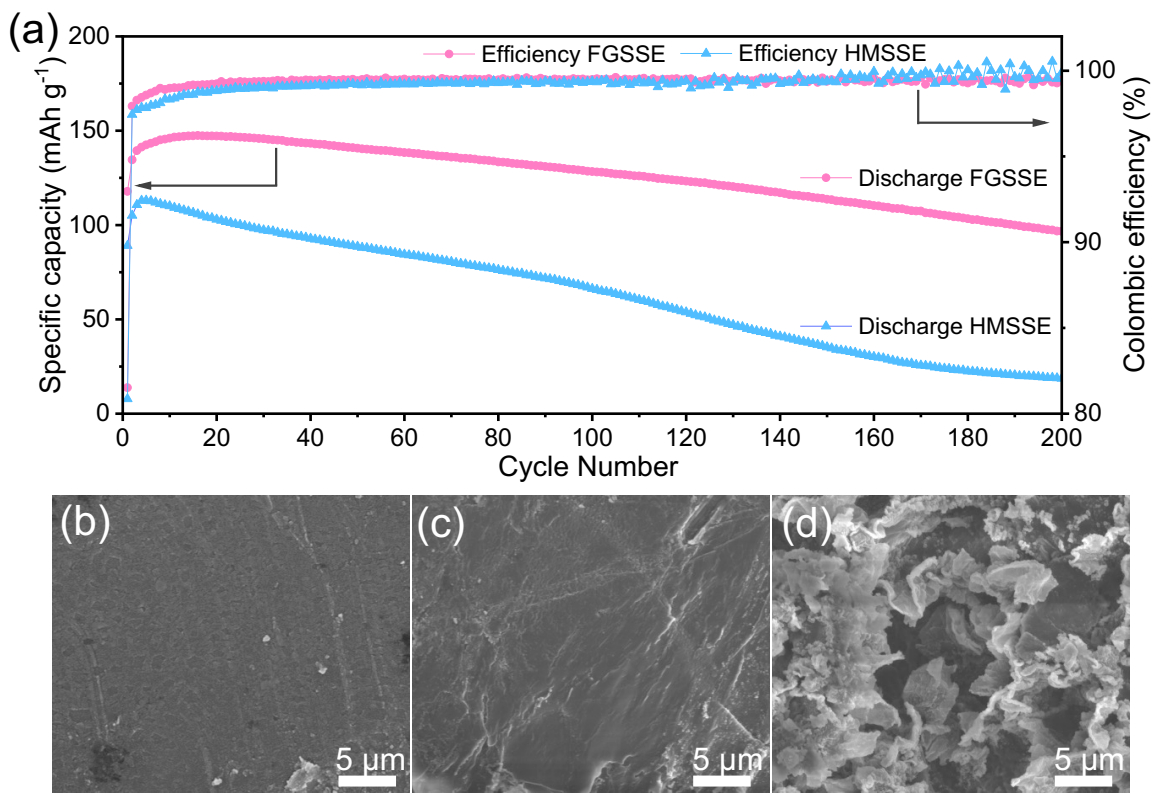

Figure 5 (a) long-term cycling performance of Li/FGSSE/NCM622 and Li/HMSSE/NCM622 cells at 1C at 60°C. (b) SEM image of the Li anode before cycling. (c) and (d) SEM images of the Li anode after 200 cycles with FGSSE and HMSSE electrolytes, respectively.

## 3. Conclusion

In summary, FGSSE with expanded electrochemical window and improved ionic conductivity is successfully prepared via an innovative CAJP method for the first time. Owing to the unique capability of modulating the ink mixing ratio on the fly, FGSSE is printed with oxidation tolerant PAN deposited in contact with the cathode followed by a gradual and smooth transition to reduction-resistant PEO. In this way, the side reaction among the electrodes and electrolyte can be avoided. And the electrochemical window of the electrolyte is expanded to 0–5.5 V. Besides that, the continuous compositional gradation of the electrolyte within the FGSSE can improve the conductivity by avoiding the sharp compositional changes and the associated interfacial barrier for ion transport. Thus, the overall conductivity of FGSSE is over 7 times of that of HMSSE at 25 °C.

The EIS analysis reveals the resistance from the electrolyte/electrolyte interface in the HMSSE is 5.7 times of the total resistance of FGSSE. By coupling with lithium anode and NCM622 cathode, the cell with FGSSE displays a capacity of 142 mAh g$^{-1}$ at 1 C and can stably run for more than 200 cycles, along with an improved rate performance. This innovative CAJP opens new opportunities to produce functionally graded SSE with enhanced properties and facilitates the development of high-performance solid-state batteries.

**Acknowledgements**

This work was supported by the National Science Foundation under awards CMMI-1747685 and CBET-2044386. The authors gratefully acknowledge the Notre Dame Center for Environmental Science and Technology and Notre Dame Integrated Imaging Facility for providing characterization support. The authors also thank the ND Energy Materials Characterization Facility (MCF) for the use of the Raman Microscope and XRD. The MCF is funded by the Sustainable Energy Initiative (SEI), which is part of the Center for Sustainable Energy at Notre Dame (ND Energy). The authors thank Yuxuan Liao and Ali Newaz Mohammad Tanvir, and Kaidong Song for their assistance throughout the study.

**Combinatorial Printing of Functionally Graded Solid-State Electrolyte for High-Voltage Lithium Metal Batteries**

*Qiang Jiang, Stephanie Atampugre, Yipu Du, Lingyu Yang, Jennifer L. Schaefer *,*

*Yanliang Zhang **

Department of Aerospace and Mechanical Engineering, University of Notre Dame, Notre Dame, IN 46556, USA

Department of Chemical and Biomolecular Engineering, University of Notre Dame, Notre Dame, IN 46556, USA

**Experimental Section**

*Materials*: Poly(acrylonitrile) (PAN, Mw = 150,000 g mol$^{-1}$), poly(ethylene oxide) (PEO, Mw = 100,000 g mol$^{-1}$), dimethylformamide (DMF), N-methylpyrrolidone (NMP), succinonitrile (SN,) were obtained from Sigma Aldrich. Lithium bis(trifluoromethanesulfonyl)imide (LiTFSI) was obtained from TCI chemicals. LLZTO (300 nm) was obtained from Neware. Lithium chips, LiNi$_{0.6}$Mn$_{0.2}$Co$_{0.2}$O$_2$ (NCM622), and Super P were obtained from MSE supplies. All chemicals were used without further purification.

*Characterization*: X-ray diffraction (XRD) patterns were performed by D8 Discover, Bruker (40 kV, 40 mA) with the scan angle from 10 ° to 60 °. The morphologies and energy-dispersive spectral (EDS) images of the SPEs were investigated with a scanning electron microscope (SEM) Helios G4 UX. Fourier transform infrared spectroscopy (FT-IR) spectra was obtained in a Bruker Tenor 27 from 4000 to 500 cm$^{-1}$ with a diamond lens attenuated total reflectance (ATR) module. The spatial distribution of organic functionalities in the electrolyte was checked by a Raman microscope (NRS-5100, Jasco). Ionic conductivity measurements were conducted on a Novocontrol Broadband Dielectric spectrometer equipped with an alpha-A high performance frequency analyzer and Quatro temperature control system with a cryostat. Data was collected in a frequency range from $1 \times 10^6$ Hz to 0.1 Hz at an AC voltage amplitude of 0.1 V from 25 to 85 °C at intervals of 15 °C. The temperature was ramped at 5 °C/min with 5 min of stabilization time at each measurement temperature. Cycling tests of the coin cell were conducted by a Neware battery test system with a voltage range of 2.8-4.3 V. Cyclic voltammetry (CV) was conducted using a Parstat, AMETEK potentiostat/galvanostat with a scanning rate of 0.1 mV s$^{-1}$ and a voltage range of 2.8-4.3 V. Electrochemical impedance spectroscopy (EIS) measurement was conducted using a Gamry Interface 1010E Potentiostat with frequency range from $1 \times 10^6$ Hz to 0.1 Hz at an AC voltage amplitude of 0.1 V. Linear sweep voltammetry (LSV) measurements were conducted using a Parstat, AMETEK potentiostat/galvanostat with a scanning rate of 0.2 mV s$^{-1}$. All the electrochemical tests were conducted at 60 °C.

*Preparation of Inks*: For PEO-based ink, 20 mg of PEO, 20 mg of LLZTO, and 10 mg of LiTFSI were added into a glass vial with 2 ml DMF. The glass vial was sonicated for 1 h in a bath sonicator before use. For PAN-based ink, it follows the same procedure as the preparation of PEO ink with the following changes: 20 mg of PAN was used to replace the PEO, 5 mg SN was added as plasticizer, and 2 ml solvent mixture (NMP: DMF=1:2) was used to improve the printability of the PAN ink.

*Printing of electrolyte films*: For the gradient structure electrolyte films, 6 layers of pure PAN ink were deposited onto the cathode discs. Then 10 intermediate layers were deposited. When printing the intermediate layer, the ink ratio of PAN:PEO changed from 1:0 to 0:1 gradually. For each layer, the ratio between PAN ink and PEO ink was fixed which means that the composition of each layer was the same. After that, 6 layers of pure PEO ink were deposited onto the top. For the HMSSE film, 11 layers of pure PAN ink were printed on the cathode discs followed by 11 layers of pure PEO ink. All the electrolyte films were dried for 48 hours in a vacuum oven at 65 °C to remove the solvent.

*Electrode Preparation and cell Assembly*: The cathode was prepared by casting onto a carbon coated aluminum current collector from a NMP slurry. The NMP slurry was prepared by mixing and stirring the NCM622 active material (70%), conductive carbon (Super P) (15%), PVDF binder (15%) and NMP in a vial overnight. The electrode was vacuum dried at 120 °C for 24 h. The mass loading of the cathode material was ~2 mg cm$^{-2}$. Then, the cathode was cut into small squares with length of ~6 mm and cold pressed at ~10 MPa. The obtained cathodes were used as the substrate during the printing of electrolyte. 2032 coin cells were assembled in an argon-filled glovebox with Li metal anode, NMC622 cathode, and the in-situ printed solid-state electrolyte.

*Measurement and calculation of Li$^+$ transference number*: In order to measure the Li$^+$ transference number, solid electrolytes were printed onto thin copper film (~6 × 6 mm). After printing, the thin copper films with electrolyte were dried in a vacuum oven at 65 °C for 48 h. Then, the Cu/electrolyte/Li cell was obtained by assembling the copper film with lithium metal into 2032-coin cells. With this asymmetric cell, Li was then plated onto the Cu electrode with current of 0.2 mA cm$^{-2}$ for 2.5h to create the Li/electrolyte/Li configuration in-situ. After resting for 10 h, EIS and a DC polarization (10 mV) experiment were performed. The Li$^+$ transference number can be deduced from the Bruce–Vincent–Evans equation as follows,

$$t_{Li^+} = \frac{I_{ss}(\Delta V - I_0 R_0)}{I_0(\Delta V - I_{ss} R_{ss})},$$

whereas *ΔV* (10 mV) is the voltage applied to the cell, $I_0$ and $R_0$ are the initial current and impedance of the cell before polarization, respectively, and $I_{ss}$ and $R_{ss}$ are the steady-state current and impedance of the cell after the polarization, respectively.

Table S1. Aerosol jet printing parameters of the electrolyte

| Parameters | Values |
| --- | --- |
| Nozzle diameter (μm) | 233 |
| Nozzle diameter (μm) | 20 |
| Sheath gas flow rate (sccm) | 60 |
| Platen temperature (°C) | 65 |
| Print speed (mm/s) | 4 |

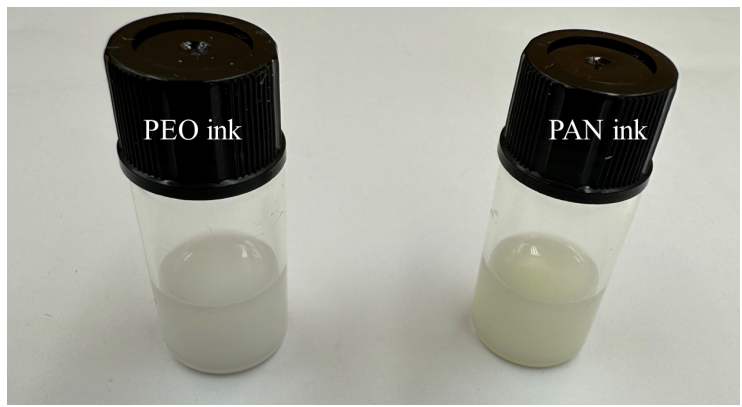

Figure S1. Image of the PEO and PAN inks.

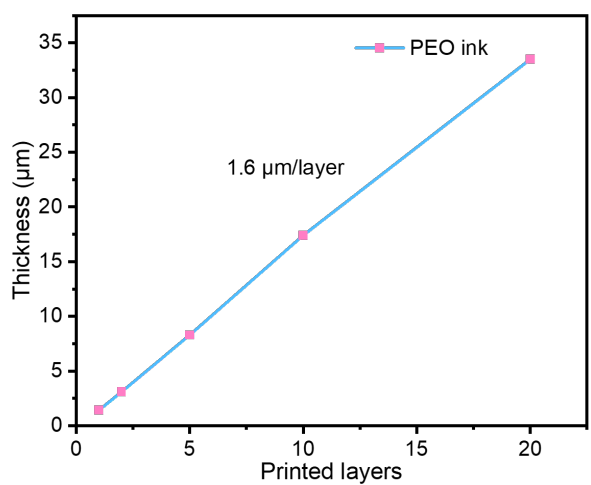

Figure S2. The relationship between the printed layer number and electrolyte film thickness.

The liner relationship between the printed layer number and electrolyte film thickness is shown in Figure S2. The average thickness of one layer was ~1.6 µm. The total thickness of the printed electrolyte film can be precisely controlled by the printed layer number.

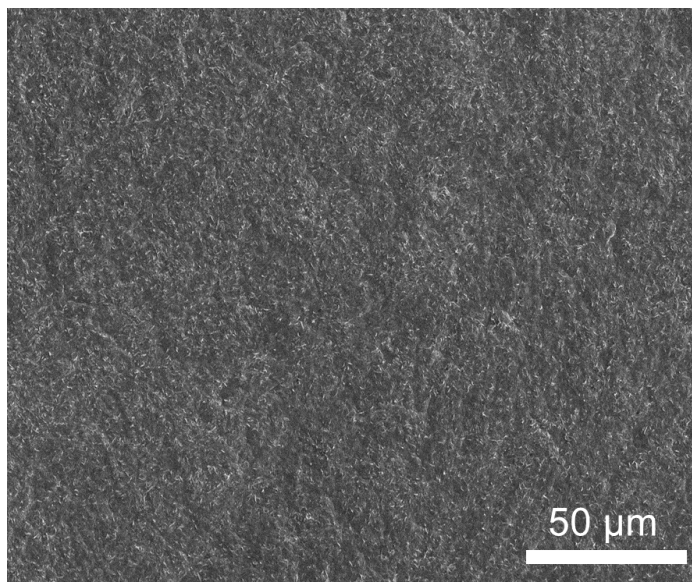

Figure S3. The SEM image of the surface of gradient electrolyte film.

Smooth and crack-free surface of the gradient electrolyte film can be obtained after optimizing the printing process which can afford the good contact between the electrode and the electrolyte.

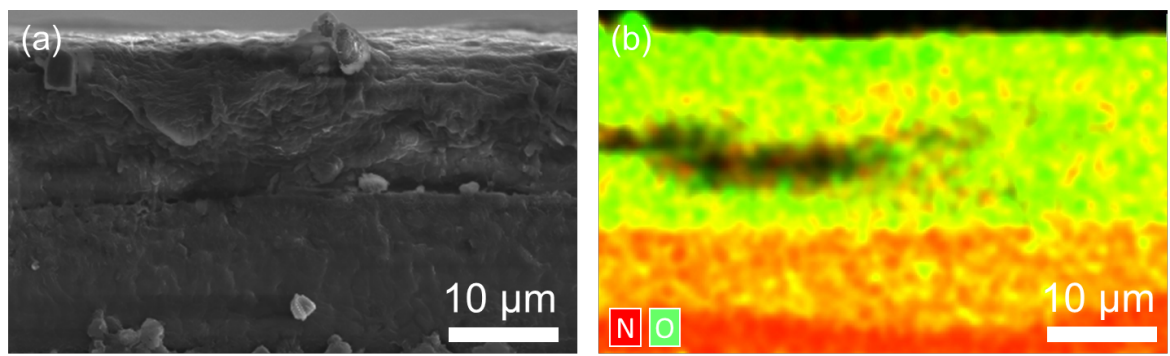

Figure S4. (a) and (b) The SEM and EDS image of the HMSSE film.

The EDS N and O distributions show that there is obvious composition change and this will lead to the obvious interface between polymers causing the discontinuous transport of ions in the film.

It should be noted that the diffusion of the PEO ink led to some PEO enter into the PAN film. But this process still has limited effect on diminishing the interface.

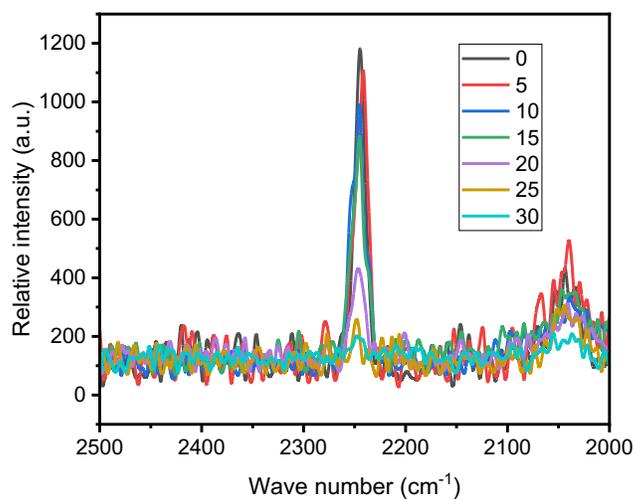

Figure S5. Raman spectra of the gradient electrolyte film at different locations.

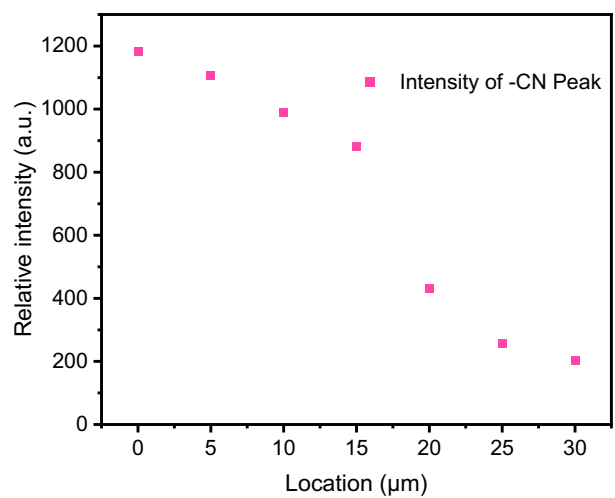

Figure S6. The relative intensity of Raman peak at 2244.7 cm$^{-1}$ of gradient electrolyte film at different locations.

The peak at 2244.7 cm$^{-1}$ belongs to the C≡N bond. It doesn't overlap with other bonds, and therefore its intensity was used to the estimate relative content of PAN to understand the spatial distribution of PAN. The intensity at 2244.7 cm$^{-1}$ of the spectrum at 30 μm is 201.9, which is similar to the base line (Figure S6). So, this value was taken as baseline when normalizing the intensity at different locations.

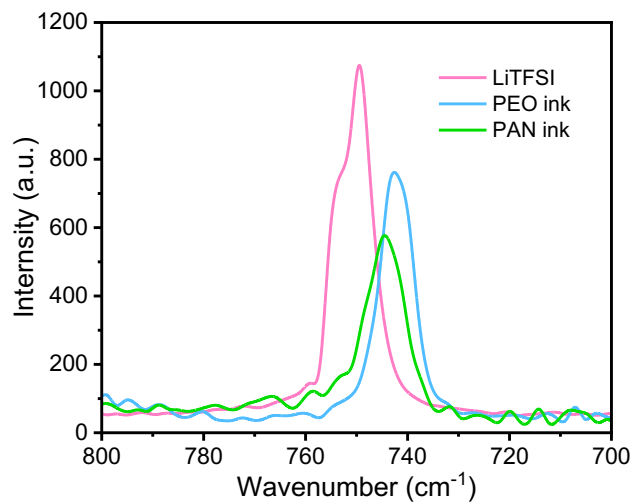

Figure S7. Raman spectra for LiTFSI salt and PEO-based and PAN-based inks.

Raman spectra for LiTFSI salt and PEO-based and PAN-based inks are shown in Figure S7. The TFSI⁻ peak at 749.5 cm⁻¹ from LiTFSI is very sensitive to the complex environment and shifts to 744.5 cm⁻¹ and 742.6 cm⁻¹ for the TFSI⁻ in PAN and PEO electrolytes.[1]

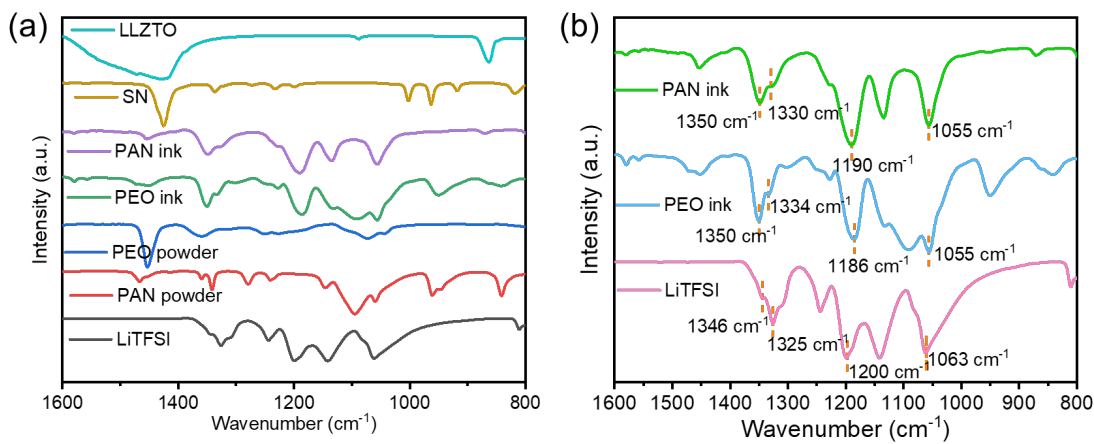

Figure S8. The FT-IR spectra of ink components (a) and (b).

FTIR shows the vibrational shifts of LiTFSI in the electrolyte. The asymmetric S-N-S stretching [$v_a$(SNS)] at 1063 cm$^{-1}$ shifts to 1057 cm$^{-1}$. The $v_a$(CF$_3$) shifts from 1200 cm$^{-1}$ to 1186–1190 cm$^{-1}$.

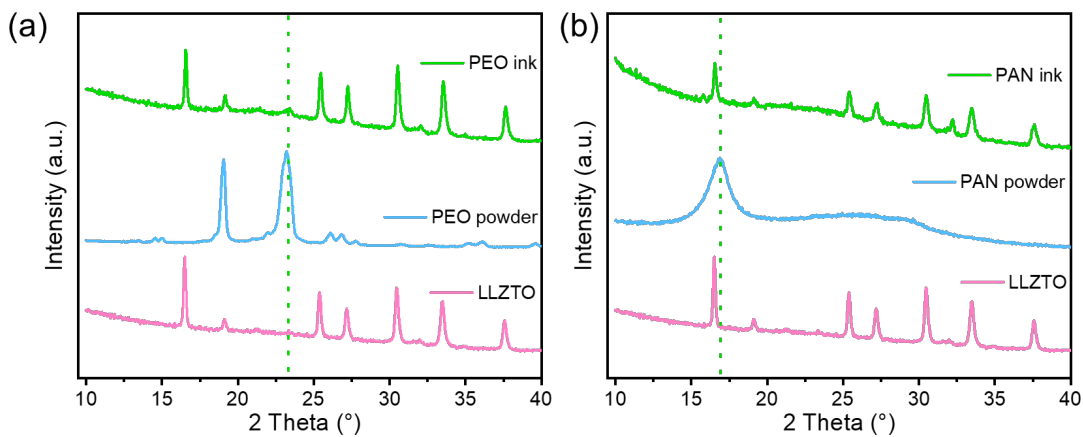

Figure S9. (a) XRD patterns of PEO, PEO ink, and LLZTO nanoparticles. (b) XRD patterns of PAN, PAN ink, and LLZTO nanoparticles.

The XRD measurement manifested the interaction between polymer electrolytes and the LLZTO nanoparticles. The disappearance of characteristic peaks (19.5° and 23.5° for PEO and 16.9° for PAN) after the introduce of LLZTO nanoparticles indicates that the incorporation of LLZTO nanoparticles efficaciously decreases the crystallinity of the PEO and PAN matrix, and greatly enhance the movement of the polymer chain which is beneficial to the ion transport.[2,3]

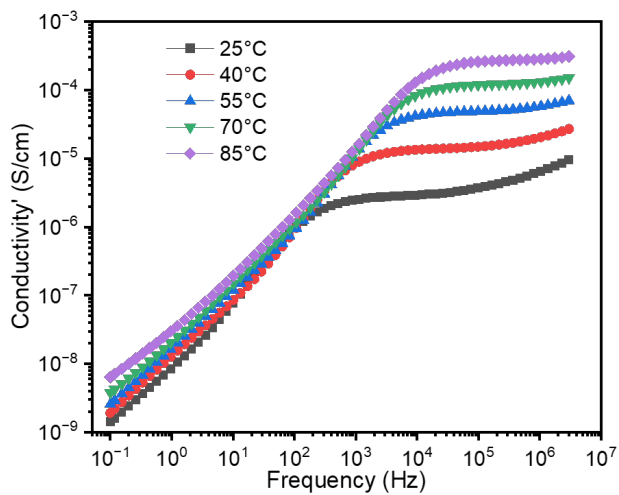

Figure S10. Real conductivity (conductivity') as a function of frequency and temperature for HMSSE. The value at the plateau was extracted as the bulk conductivity.

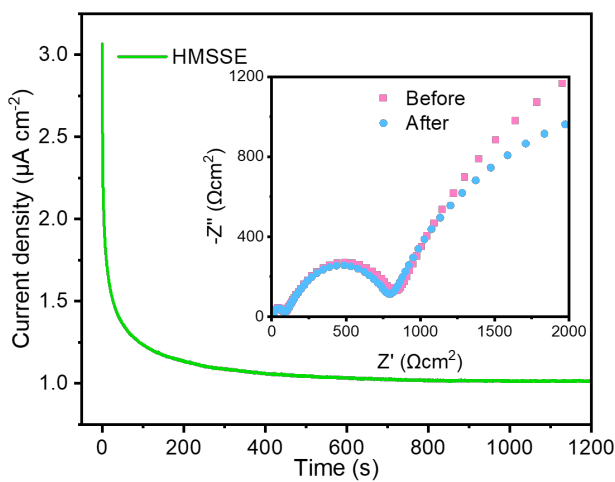

Figure S11. Impedance spectra and DC polarization curve of Li/HMSSE/Li for Li$^+$ transference number tests.

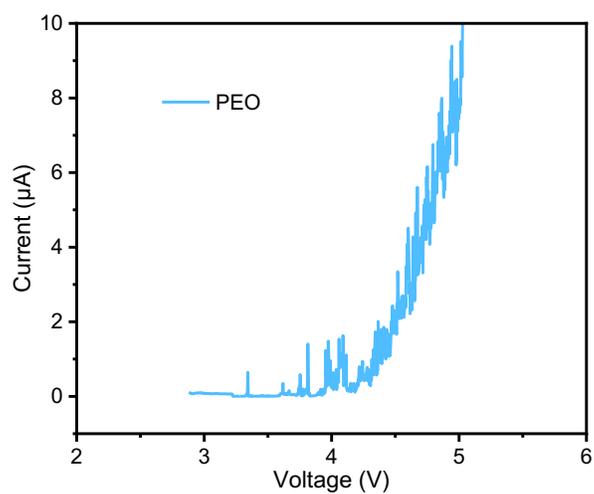

Figure S12. Linear scan voltammogram of Li/PEO electrolyte/SS to determine oxidative stability.

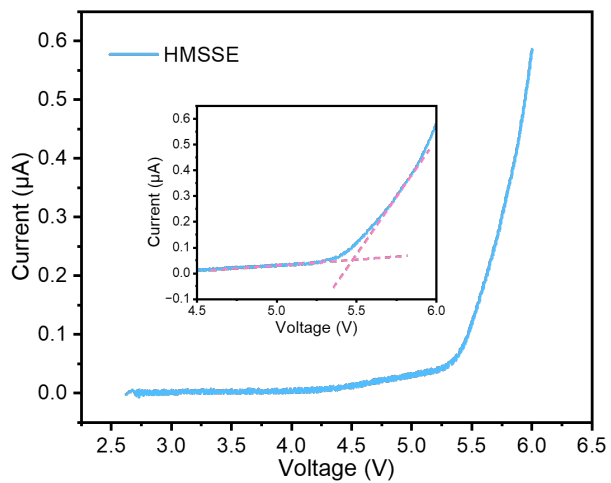

Figure S13. Linear scan voltammogram of Li/HMSSE/SS to determine oxidative stability. Inset is the zoom-in image of the onset of the oxidation process.